\documentclass{aa}  

\usepackage{graphicx}
\usepackage{txfonts}
\usepackage{multirow}

\usepackage[usenames,dvipsnames]{color}
\usepackage{xspace}
\usepackage{multicol}

\newcommand{\Vcritfrac}{${\varv_{\textrm{rot}}}/{\varv_{\textrm{crit}}}\thinspace$}
\newcommand{\V}{$\varv_{\textrm{rot}}$\xspace}
\newcommand{\Vcrit}{$\varv_{\textrm{crit}}$\xspace}
\newcommand{\Msun}{\ensuremath{\,\mathrm{M_\odot}}\xspace}

\newcommand{\Myr}{\,Myr\xspace}

\newcommand{\kms}{\,km\,s$^{-1}$\xspace}

\newcommand{\GG}[1]{}

 \def\simle{\mathrel{\hbox{\rlap{\hbox{\lower4pt\hbox{$\sim$}}}\hbox{$<$}}}}
 \def\simgr{\mathrel{\hbox{\rlap{\hbox{\lower4pt\hbox{$\sim$}}}\hbox{$>$}}}}

\graphicspath{%
  {./figs/pdfs/}%
  }

\usepackage{natbib,twoopt}
\usepackage[breaklinks=true]{hyperref} 

\begin{document}

   \title{The Single Star Path to Be Stars}

   \author{B. Hastings 
            \and
            Chen Wang
            \and
            N. Langer
            }
    \institute{Argelander-Institut f\"{u}r Astronomie, Universit\"{a}t Bonn, Auf dem H\"{u}gel 71, 53121 Bonn, Germany\\ \email{bhastings@astro.uni-bonn.de}
                \\
             }
             
    \authorrunning{Hastings, Wang, and Langer}         

 
\abstract
   {Be stars are rapidly rotating B main sequence stars, which show line emission due to an outflowing disc. 
 By studying the evolution of rotating single star models, we can assess their contribution to the observed Be star populations.} 
   {We identify the main effects which are responsible for single stars to approach critical rotation as functions of initial mass 
   and metallicity, and predict the properties of populations of rotating single stars. }
   {We perform population synthesis with single star models of initial masses ranging between 3 and 30 \Msun, 
  initial equatorial rotation velocities between 0 and 600\kms at compositions representing the Milky Way, 
  Large and Small Magellanic Clouds. These models include efficient core-envelope coupling mediated by internal magnetic fields 
  and correspond to the maximum efficiency of Be star production. We predict Be star fractions and the positions of fast rotating stars 
  in the colour-magnitude diagram.}
   {We identify stellar wind mass-loss and the convective core mass fraction as the key parameters which determine the 
   time dependance of the stellar rotation rates. 
   Using empirical distributions of initial rotational velocities,
   our single star models can reproduce the trends observed in Be star fractions with mass and metallicity. However,
   they fail to produce a significant number of stars rotating very close to critical. 
   We also find that rapidly rotating Be stars in the Magellanic Clouds should have significant surface nitrogen enrichments, 
   which may be in conflict with abundance determinations of Be stars.}
  {Single star evolution may explain the high number of Be stars if 70 to 80\% of critical rotation
  would be sufficient to produce the Be phenomenon. However even in this case, the unexplained presence of many Be stars far below the cluster turn-off indicates the importance of the binary channel for Be star production.}


   \maketitle
%

\section{Introduction \label{sec:introduction}}

Ever since their discovery over 150 years ago \citep{BeDiscovery}, Be stars have offered a promising, although misted window into massive star evolution and structure. It was proposed by \citet{Struve} that Be stars are fast rotators, whose emission lines originate from a circumstellar decretion disc, a picture which is maintained until today \citep{2013A&ARv..21...69R}. Yet, it is still not clear how fast a B-type star must rotate in order to become a Be star. 

For a decretion disc to form, the equatorial rotation velocity \V  is expected to be a significant fraction of the critical rotation velocity, $v_{\textrm{crit}}$, defined as the rotation velocity at which material at the equator becomes unbound from the star. Observational evidence suggests that the threshold rotation rate for the Be phenomenon is mass dependant, and could be as low as \Vcritfrac = 0.6 for stars more massive than 8.6 \Msun and as high as \Vcritfrac = 0.96 for stars with $M< 4$ \Msun \citep{HuangGiesObs}. Similarly \citet{2016A&A...595A.132Z} find that the Be phenomenon is characterized by a wide range of true velocity ratios (0.3 $<$ \Vcritfrac $<$ 0.95) and that the probability that Be stars are critical rotators is small. In this case one must look for an additional mechanism to feed the Be disc. Pulsations seem promising as they can serve to kick matter from the surface of a star, however it is found that not all Be stars pulsate \citep{2002A&A...383L..31B} and that among those that do there is a wide range of pulation frequencies and types \citep{2013A&ARv..21...69R}. Another possibility is that the disc is fed through outbursts of magnetically active starspots, similar to coronal mass ejections as seen in the Sun, as suggested by \citet{2019arXiv191103068B} based on recent TESS results.

On the other hand, \citet{Townsend} have argued that all Be stars in fact rotate very close (\Vcritfrac $>0.95$) to the critical velocity, with those which have low measured rotation rates being strongly affected by gravity darkening. Following the Von Zeipel law \citep{1924MNRAS..84..665V}, gravity darkening in a fast rotating star makes the stellar pole, which has a low rotational velocity, more luminous than the equator which has a high rotational velocity, resulting in the star appearing as though it is rotating slower than in reality. 

A further question surrounding Be stars is why the phenomenon seems to be restricted mostly to B-type stars and why Be stars are more common in certain spectral classes than others. Observations in the Milky Way show that the fraction of Be stars in a certain spectral class varies across spectral type with the most Be-stars found at B1-B2 classification, where the Be fraction is $34\%$ while in comparison the Be fraction for B9 stars is $8\%$ and the total fraction of Be stars to B stars was measured to be $17\%$ \citep{BeFrequencies}. Furthermore, Oe stars seem to be rather rare, with less than 20 having been detected in the Milky Way \citep{Oe}. It is not clear whether this is caused by processes within O stars themselves, the mechanisms responsible for forming a disc around a fast rotating star, or the conditions under which very massive stars form. 

The existence of around 150 \citep{BeXRBcat} detected binary systems consisting of a Be star and a compact object (so called Be/X-ray binaries) demonstrates that binary interactions can spin up a star significantly \citep{1975BAICz..26...65K,PolsBinaryModels,2006A&A...455.1165L,2012ARA&A..50..107L}. In order to build a full model of the Be phenomenon, binary evolution must be as well understood as the channels for forming single Be stars. 

 \citet{1997A&A...322..116V} used binary star evolution calculations to predict that at most 20\% of the Be star population results from binary interactions. Despite this, observations of surface nitrogen abundances in Magellanic Cloud Be stars are in disagreement with fast rotating single star models \citep{BeN1}. Thus questions such as, which is the dominant Be star formation channel and what are the differences between the single and binary Be population remain open. 

Considerable efforts have been made to predict the relative fractions of Be stars through single star modelling, most notably by \citet{EkstromBe}. These models include coupling between core and envelope by hydrodynamic viscosities through the shear instability. However, a stronger coupling may be produced by internal magnetic fields known as the Tayler-Spruit dynamo \citep{TSdynamo}  which produces very efficient angular momentum transport throughout the star so that near solid-body rotation occurs. 
The models analysed here include such effects, and therefore employ the maximum efficiency of
spinning up the surface layers due to the core contraction during hydrogen burning, which is a key factor to 
produce single Be stars.  
At present the Tayler-Spruit dynamo is used to explain relatively slow rotation rates in white dwarfs \citep{WDslow} and young pulsars \citep{Pulsarslow}.

In Section \ref{sec:method} the models are introduced and our approaches are outlined. In Section \ref{sec:results}, models from the grid are analysed under conditions of fast and slow rotation for MW and SMC metallicities, the factors governing the approach to the critical velocity are investigated and the expected surface nitrogen abundances of Be stars is investigated. In Section \ref{sec:popsyn}, population synthesis is performed to calculate the expected fractions of single Be stars in clusters of differing ages and metallicities and predict the positions of fast rotating stars in the colour-magnitude diagram.

\section{Method \label{sec:method}}

\subsection{Stellar Models \label{sec:modelsesc}}
We analyse the single star evolutionary models of \citet{Brott1} to predict the properties of rotating single stars 
throughout their main-sequence evolution. We consider masses from 3\Msun to 30\Msun at various initial rotation velocities 
ranging from zero to approximately 600\kms. Because the model grid is spaced in initial rotational velocities,
but the initial critical rotation velocity increases with mass, our grid does not contain models
with initial values of \Vcritfrac greater than around 0.7 for initial masses greater than 25\Msun.  

Three initial chemical compositions represent metallicities of the Milky Way (MW), Large Magellanic Could (LMC) and Small Magellanic Cloud (SMC). The models include internal transport of angular momentum via the Taylor-Spruit dynamo \citep{TSdynamo} which has the effect of enforcing near solid-body rotation throughout most of the main-sequence evolution. 
The adopted mass-loss scheme is given by \citet{VinkMassloss}. An enhancement of the mass loss due to rotation is used as outlined in \citet{2005A&A...443..643Y}, whereby the mass loss rates are increased by a factor depending on the ratio of the rotation velocity to the critical velocity like

\begin{align}
\label{Eq:massloss}
\dot{M}(\Omega) = \dot{M}(0) \Bigg(\frac{1}{1-\frac{\varv_{\textrm{rot}}}{\varv_{\textrm{crit}}}} \Bigg) ^{0.43}
\end{align}
where 
\begin{align}
\varv_{\textrm{crit}} =\sqrt{\frac{GM}{R}(1-\Gamma)} \ ; \  \Gamma=\frac{\kappa L}{4 \pi c G M} . \label{Eq:Vcrit}
\end{align}

For a detailed description of the models see \citet{Brott1}.

\subsection{Population Synthesis \label{sec:popsynmeth}}
To predict properties of populations of rotating stars we use population synthesis to model open star clusters (i.e. collections of coeval stars without any continuous star formation) at various ages. For a cluster age $t$, we select pairs of random values from a Salpeter initial mass distribution (with exponent 2.35) and a distribution of initial critical velocity fraction, $M_i$ and \Vcritfrac$_i$. Then we find the masses, $M_1, M_2$ on the model grid that are straddling the chosen mass value, such that $M_1<M_i<M_2$. For $M_1$ and $M_2$ we interpolate the hydrogen burning lifetimes, $t_{MS}$ as a function of initial critical velocity fraction to obtain the hydrogen burning lifetimes at the chosen value, \Vcritfrac$_i$. Next the hydrogen burning lifetime, $t_{MS,i}$, of a model with mass $M_i$ and initial rotation \Vcritfrac$_i$ is found by interpolating between the hydrogen burning lifetimes of $M_1$ and $M_2$. The fractional lifetime is then given by ${t}/{t_{MS,i}}$. If the fractional lifetime is greater than 1, the star will not be hydrogen burning anymore so the process is abandoned and new samples are drawn. We then select models with masses $M_1$ and $M_2$ at fractional hydrogen burning times ${t}/{t_{MS,i}}$. An interpolation of the quantity of interest, $Q$ across initial critical velocity fraction gives the values of $Q$ for masses $M_1, M_2$ with initial rotation \Vcritfrac$_i$ and fractional hydrogen burning time  ${t}/{t_{MS,i}}$. One final interpolation between $M_1$ and $M_2$ gives the predicted quantity of the selected mass $M_i$ at the given cluster age. The quantities of interest are luminosity, critical velocity fraction at the current time and effective temperature. 

\begin{figure}
	\includegraphics[width=1.0\linewidth]{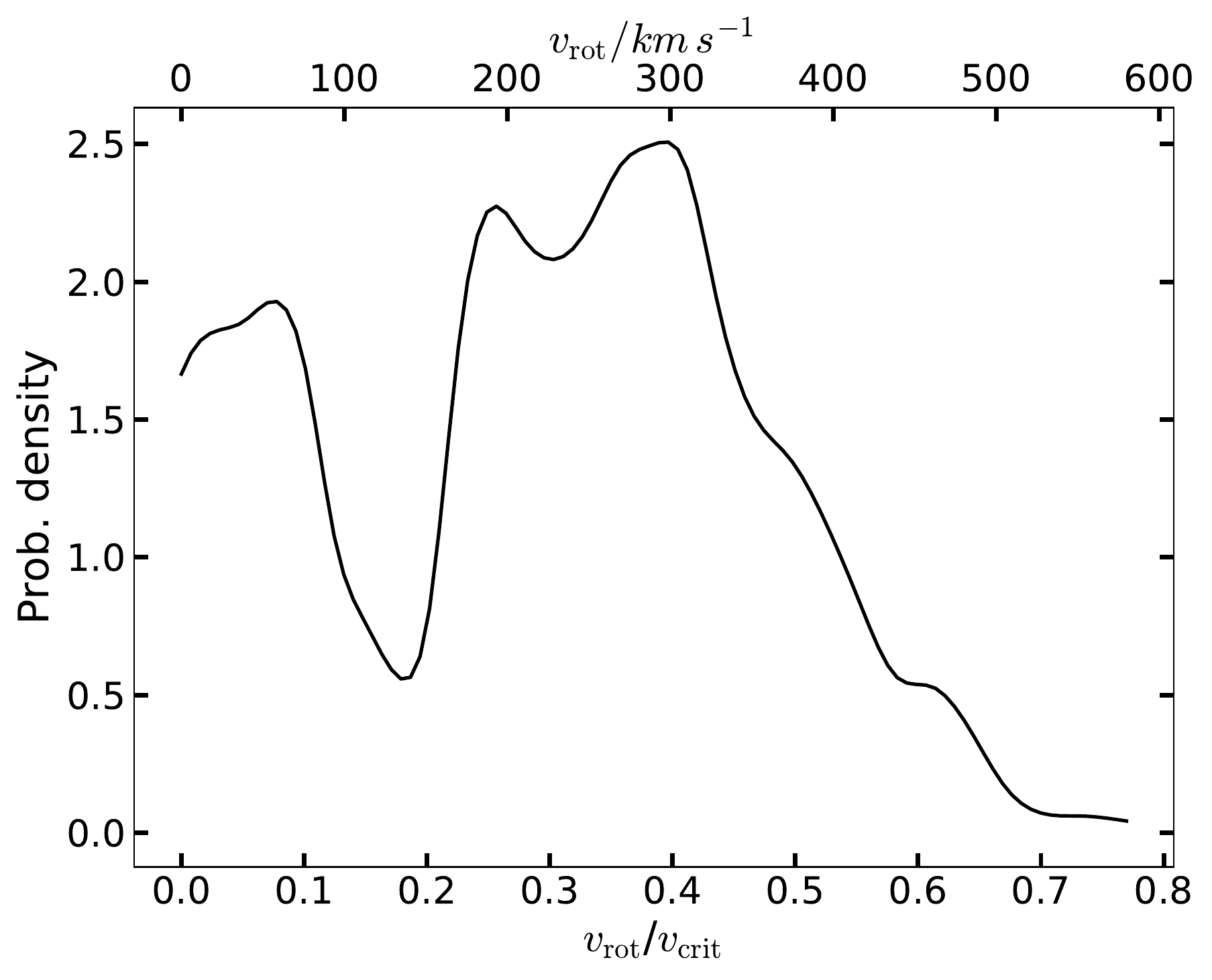}
	\centering
	\caption{The deconvolved rotation distribution of early B stars as observed by VLT-FLAMES Survey (Dufton et al., 2013). The distribution was converted to \Vcritfrac using 15\Msun LMC models at ZAMS. The upper scale shows how the critical velocity fraction values, \Vcritfrac match to the equatorial velocities, \V. }
	\label{fig:rotdist} 
\end{figure}

The initial rotational velocity distribution used was taken from VLT-FLAMES observations of early B\,stars in the 30 Doradus region of the LMC \citep{VFTSB} and is shown in Fig.\,\ref{fig:rotdist}. The deconvolved distribution of equatorial rotational velocities was converted to a distribution in critical velocity fraction by applying a mapping between the two as determined from the 15\Msun LMC models at ZAMS and then normalizing such that the integral over the whole probability density function equals unity. It is noted that for the heaviest masses on the grid, the distribution extends beyond its limits. When such a massive, very fast rotating star is chosen from the distributions, instead the fastest rotator in the grid is used. Observations in 30 Doradus show that no O stars are observed to rotate with deconvolved equatorial velocities much greater than 500\kms \citep{VFTSO,2011ApJ...743L..22D}. For a 25\Msun star to rotate at a critical velocity fraction of 0.7, it would require an equatorial rotation velocity of the order 700 kms$^{-1}$. It is thus safe to assume that O stars do not enter the ZAMS with initial critical velocity fractions much greater than 0.65, or that if they do, they spin down very quickly.

We note that adopting the observed distribution of rotational velocities of \citet{VFTSB} as the initial velocity distribution
for stars in our synthetic populations may introduce an inconsistency, since the the sample of \citet{VFTSB} consists of field 
stars of all ages. However, as discussed in \citet{VFTSB} (see also Sect.\,3.1 below), the rotational velocities of 
the considered single stars are expected to change very little during their main sequence evolution. 
If binary evolution affects this distribution \cite{2013ApJ...764..166D}, then we would overestimate
the number of stars which are born rotating very rapidly. In this case, the predicted number of Be stars from our models
may be considered as upper limits. 

To compare the models with observations of the SMC open cluster NCG 330 \citep{MiloneObs2} in the colour-magnitude diagram, the bolometric luminosities and effective temperature are converted to absolute magnitudes in the Hubble Space Telescope Wide-Field Camera 3 filters F814W and F336W by interpolating tables based on synthetic stellar spectra \citep{2002A&A...391..195G}. The values of distance modulus and reddening adopted are $(m-M)=18.92$ and $E(B-V)=0.06$ respectively. The absorption coefficients used are $A_{F814W} = 2.04E(B-V)$ and $A_{F336W} = 5.16E(B-V)$ \citep{MiloneObs}. The effects of gravity darkening are included as described by \citet{FLR}, whereby the effective temperature and luminosity of a star are multiplied by  parameters that depend on the inclination angle and fraction of angular critical velocity. Then using these corrected effective temperature and luminosity values, we calculate the absolute magnitudes as described above. The inclination angles, $i\,$, in our synthetic population are chosen such that $\cos(i)$ is  uniformly distributed between 0 and 1, meaning that it is more likely to observe any given star equator-on than pole-on. Such a distribution describes a random orientation of the rotation axis. 

\section{Results \label{sec:results}}

\subsection{Spin Evolution \label{sec: velEv}}

During the evolution of a slowly rotating star during core hydrogen burning, a strong chemical gradient develops between the convective core and the radiative envelope. The core density increases, and as a reaction the envelope must expand in order to maintain hydrostatic and thermal equilibrium. Eq.\,\ref{Eq:Vcrit} shows that as the stellar radius increases, the critical velocity decreases, thus during main-sequence evolution, the critical velocity  will fall.

In the absence of internal angular momentum transport, as core density increases, the local conservation of angular momentum will demand that the angular velocity of the core increases. Likewise as the envelope expands, the angular velocity of the envelope will decrease. This results in an angular velocity gradient developing between the convective core and radiative envelope. However when the core and envelope are coupled via angular momentum transport, angular momentum is transported from the core to the envelope, decreasing the angular velocity gradient throughout the star. The physical processes responsible for the angular momentum transport in the models studied here are magnetic torques arising from the Tayler-Spruit dynamo \citep{TSdynamo}, which leads to near solid body rotation. While the envelope is expanding and the star is rotating as a solid body, the critical rotation velocity will decrease while the equatorial rotation velocity drops only slowly or even increases
(see Section \ref{sec:coremass}). 

\begin{figure*} 
	\includegraphics[width=0.98\linewidth]{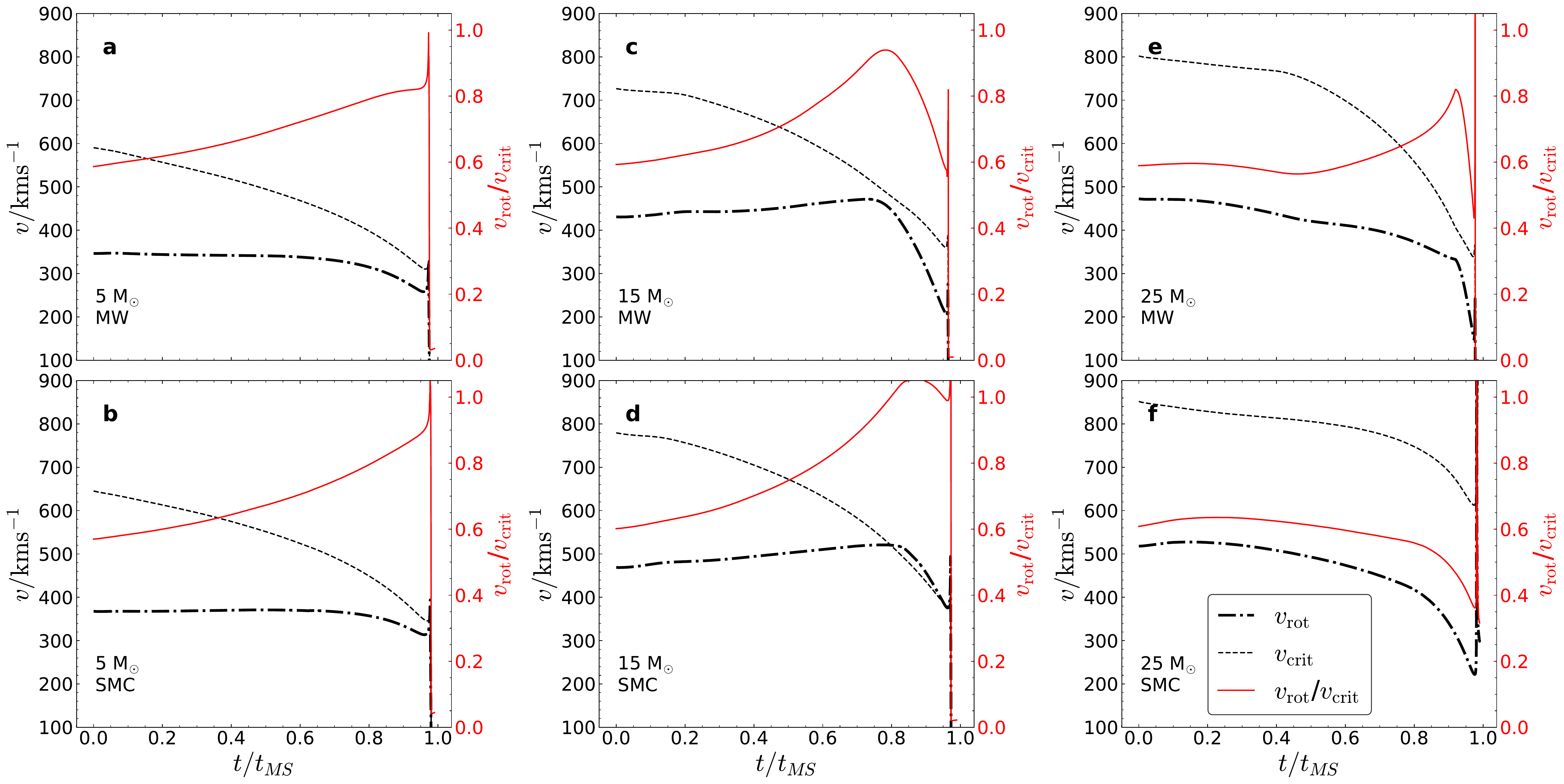}
	\centering
	\caption{The evolution of equatorial rotational velocity \V (thick dot-dashed), critical velocity \Vcrit (dashed) and the critical velocity fraction \Vcritfrac (solid red) for 5\Msun (panels a, b), 15\Msun (panels c, d) and 25\Msun (panels e, f)  models at MW metallicity (panels a, c, e) and SMC metallicity (panels b, d, f). The initial critical velocity fraction values are all approximately 0.6. The X-axis indicates the fractional main sequence lifetime, $t/t_{MS}$.}
	\label{fig:vel-ev} 
\end{figure*}

Fig.\,\ref{fig:vel-ev} gives examples of the evolution of critical velocity, equatorial velocity and fraction of critical velocity during main-sequence evolution for MW and SMC models of initial masses 5,15 and 25\Msun. All models in the plot have an initial critical velocity fraction of approximately 0.6. Although the less massive models have slower equatorial velocities, they also have lower critical velocities (because of a relatively weak dependance on radius with mass), making the critical velocity fraction nearly the same for all models in the plot. It is seen that for the 5\Msun models  the equatorial rotation velocity remains nearly constant, while the critical velocity decreases. For the 15\Msun models, the equatorial rotation velocity increases due to the effects of angular momentum transport.  As a result, the critical velocity fraction is generally increasing during hydrogen burning such that the stars are evolving closer to critical rotation, unless angular momentum is drained at a high rate due to mass loss (see Section \ref{sec:amLoss}).


\subsection{The Effect of Mass-loss \label{sec:amLoss}}
The dominating effect of mass-loss through stellar winds is to remove angular momentum, not mass. This is so because 
even in the absence of magnetic fields, the rate of angular momentum loss relative to the total angular momentum is about 10 times larger than the rate of mass loss relative to the stellar mass \citep{1998A&A...329..551L}. Thus a star's mass-loss may strongly affect the spin evolution.

Panels a, c, e of Fig.\,\mbox{\ref{fig:vel-ev}} show that models at MW metallicity experience a turn-over in the evolution of their critical velocity fraction. This is because in the late stages of hydrogen burning, the star's mass-loss rate increases significantly which has the effect of removing angular momentum from the surface at a rate which cannot be compensated by internal angular momentum transport mechanisms, meaning that solid body rotation is no longer a good approximation and the equatorial velocity decreases. This period of strong mass-loss is caused by the iron opacity bistability in which partial recombination of Fe ions at effective temperatures of around 22kK causes a sharp increase in opacity and hence mass-loss \citep{bistability}. This behaviour is strongly dependant on metallicity and so weaker in the SMC or LMC models. Comparing Panels c and d of Fig.\,\mbox{\ref{fig:vel-ev}}, it can be seen that despite both MW and SMC models starting with approximately equal critical velocity fractions, the SMC model achieves a much larger critical velocity fraction at the end of the main-sequence.  This comparison between MW and SMC models illustrates the effect of mass-loss on the approach to the critical velocity.

\begin{figure}
	\includegraphics[width=0.9\linewidth]{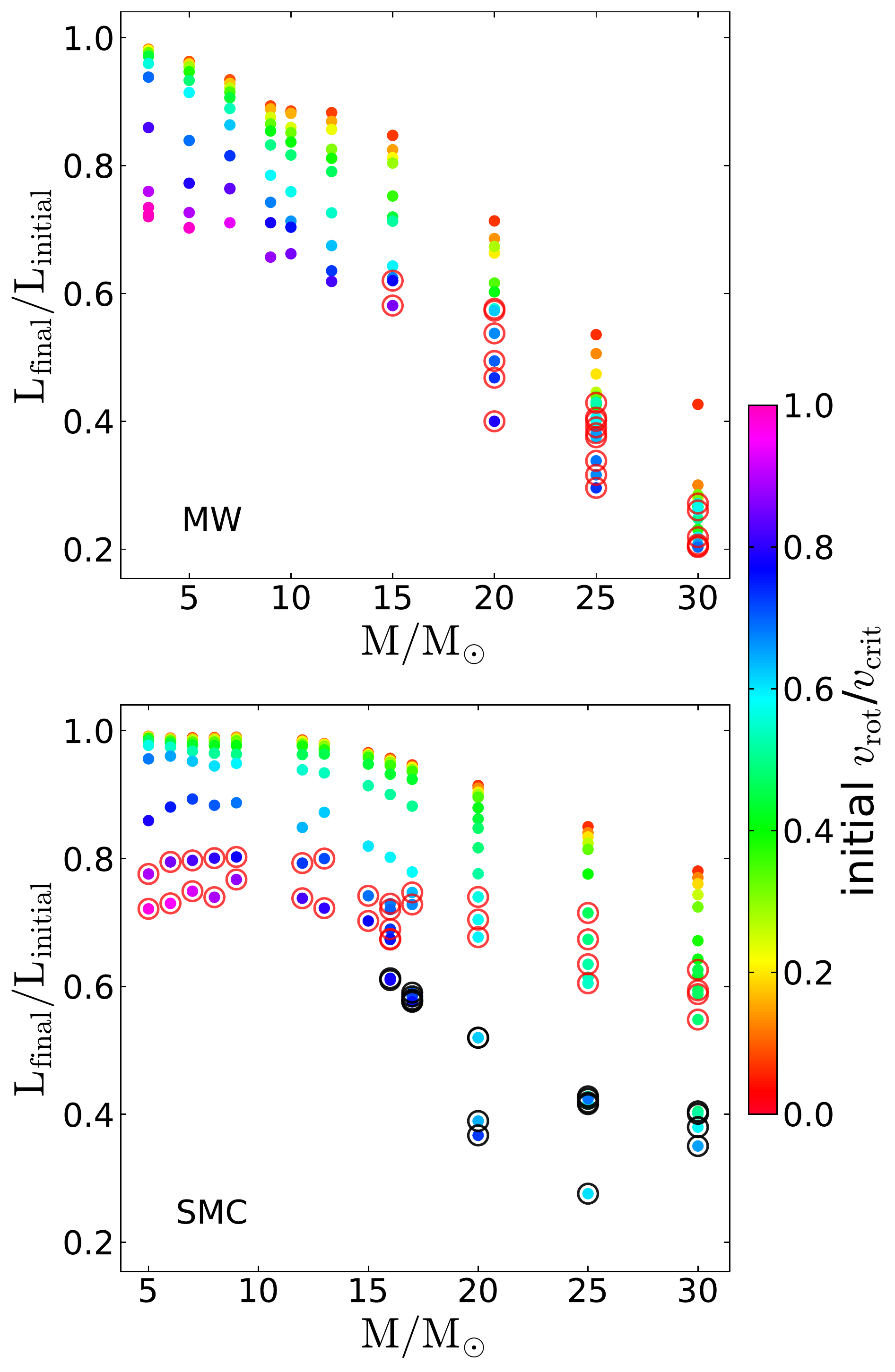}
	\centering
	\caption{The fraction of angular momentum at the end of hydrogen burning ${L_{final}}$, to angular momentum at the start of hydrogen burning ${L_{initial}}$, as a function of mass for MW  (top) and SMC (bottom) metallicities. The colour of the points represents the initial critical velocity fraction \Vcritfrac. Models with a growing helium surface abundance throughout the duration of their evolution are marked by a black circle. Models with a growing helium surface abundance for part of their evolution are marked by a red circle. }
	\label{fig:angmomloss} 
\end{figure}

Fig.\,\ref{fig:angmomloss}  shows the ratio of angular momentum at the end of hydrogen burning to the initial angular momentum for models of varying mass and initial critical velocity fraction for MW and SMC models. It is clear that almost every SMC model loses much less angular momentum than the corresponding model at MW metallicity. The exceptions are fast rotating massive SMC models which undergo quasi-chemically homogeneous evolution, and during so become very luminous which leads to increased mass loss rates. For SMC models one may judge that mass-loss becomes irrelevant to the angular-momentum budget below masses of around 10\Msun, where most models (except very fast initial rotators, say with \Vcritfrac > 0.7) retain more than 90$\%$ of their angular momentum. For MW models we see that only slowly rotating models less massive than 5\Msun retain more than 90$\%$ of their angular momentum. As expected, the effect of mass-loss is strongly metallicity dependant. It is also seen that 
below 15$\dots$20 \Msun, angular momentum loss becoming less dependant on mass.

Fig.\,\ref{fig:angmomloss} shows that for any given mass the fraction of angular momentum lost is a strong function of the initial rotation. For example a 5\Msun model at MW metallicity can lose between 2 and 30 $\%$ of its initial angular momentum. One contributing factor is the lifetime effect, whereby under the effects of rotation,  rotational mixing causes hydrogen to be mixed into the centre of the star and so hydrogen burning can continue for a longer time. For all of our models, the hydrogen burning lifetime enhancement between the non-rotating model and that with an initial rotation velocity of 600 \kms never exceeds a factor of 1.5. Thus for models losing only small fractions of their total angular momentum as slow rotators (such as low mass MW metallicity models and all SMC metallicity models), the lifetime effect cannot explain wholly the increase in angular momentum loss. 

Another effect is rotationally enhanced mass loss. As a star approaches the critical rotation velocity, material at the equator becomes less tightly bound due to the centrifugal force, thus one would expect angular momentum lost through winds to increase with rotation velocity. In the models this is governed by Eq.\,\ref{Eq:massloss}. With \Vcritfrac $=0.8$, the mass loss rates are doubled, so rotationally enhanced mass loss plays only a large role when very high critical rotation fractions are achieved. As an example the 5\Msun models at MW metallicity in Fig.\,\ref{fig:angmomloss} show that in the range of initial \Vcritfrac from 0 to 0.4, the total angular momentum lost is almost the same.  However when approaching critical rotation, the enhancement factor becomes divergent, so models rotating near the critical velocity experience tremendous mass loss. 

A further effect of rotationally induced mixing is to increase the overall mean molecular weight, $\mu$ in the star, compared to models with no rotationally induced mixing. Homologous models suggest a strong dependance on luminosity with mean molecular weight such that $L \propto \mu ^{4}$ \citep{KippenhahnBook}. In turn, mass-loss rates are dependant on the luminosity, for the wind prescription used in the models the dependance is approximately $\dot{M} \propto L^2$ \citep{VinkMassloss}. Thus rotationally induced mixing leads to higher mass loss and angular momentum loss. For models which experience quasi-chemical homogeneous evolution, where the star can become a helium star, this effect becomes very apparent. Models which undergo quasi-chemical homogeneous evolution for the duration of hydrogen burning (defined by a monotonically increasing surface helium mass fraction) are marked with black circles in Fig.\,\ref{fig:angmomloss}. Similarly models which undergo a phase of quasi-chemical homogeneous evolution (defined by having a slowly increasing difference between surface and central helium mass fraction for longer than one third of the hydrogen burning lifetime) are marked by a red circle. Although these models do not have high initial critical rotation fractions, they still lose large fractions of their angular momentum. Quasi-homogeneous evolution occurs more readily in the lower metallicity models because mass loss being a strong function of metallicity, so the MW metallicity models slow down relatively quickly, rotational mixing becomes less effective and the homogeneous evolution stops \citep{Yoon}.

\subsection{The Effect of Convective Core Mass \label{sec:coremass}}

During the main-sequence evolution of a massive star, the convective core contracts while the radiative envelope expands. The conservation of angular momentum will therefore demand that in the absence of any internal angular momentum transport, the convective core and radiative envelope increases and decreases respectively their overall angular velocity (i.e. the core "spins up" while the envelope "spins down"). This tells us that to enforce solid body rotation during core contraction and envelope expansion, angular momentum must be transported from the core to the envelope. This is achieved by magnetic interactions which transport angular momentum along the angular velocity gradient within the star. 

Let us now consider two extreme examples. In a rotating star with a negligible envelope mass, the core will dominate the angular momentum budget. Thus to maintain a constant rotational velocity during envelope expansion, a relatively low angular momentum transport rate is required. On the other hand, for an envelope dominated star to rotate at a constant velocity while the envelope is expanding, the angular momentum transport rate from core to envelope must be high. It is then likely that internal angular momentum transport mechanisms are unable to meet this demand, and as a result the rotational velocity of the stellar surface will decrease due to the effect of local angular momentum conservation. 

\begin{figure*}
	\includegraphics[width=0.99\linewidth]{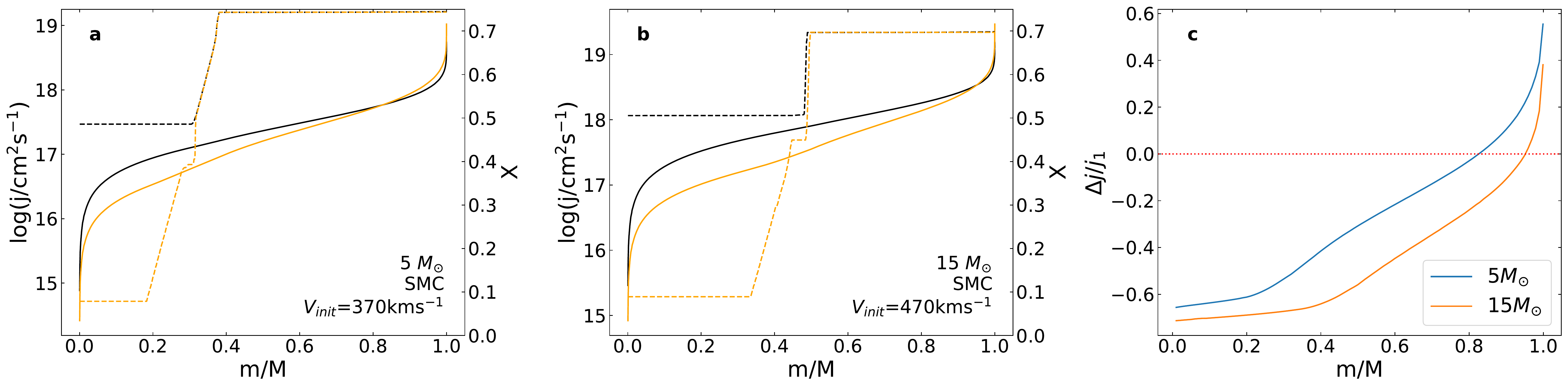}
	\centering
	\caption{\textbf{a-b:} The specific angular momentum (solid lines) and hydrogen mass fraction (dotted lines) profiles for two SMC models of masses 5 and 15\Msun and initial equatorial velocities of 370 and 470 kms$^{-1}$ respectively. These are the same models plotted in panels b \& d of Fig.\,\ref{fig:vel-ev} and both models have initial critical velocity fractions of around 0.6. Profiles are plotted for models where central helium mass fraction is 0.45 (black) , and 0.91 (orange).
\textbf{c:}	 For each model in panel a \& b the fractional difference in specific angular momentum is plotted between the two times. The blue line represents the 5\Msun model,the orange the 15\Msun model. The dotted red line gives a reference for no angular momentum transport.
The X-axis in all plots indicates the fractional mass co-ordinate. }
	\label{fig:angmomtran} 
\end{figure*}

In our models, there is an inner region of the star from which angular momentum is being transported and there is an outer region which the angular momentum is being transported to. Inbetween these regions there must be a point which neither gains nor loses any specific angular momentum. The location of this angular momentum "valve" will give an indication as to the strength of the core mass effect as discussed above. Figs.\,\ref{fig:angmomtran}. a, b show the specific angular momentum profiles of a 5 and 15\Msun model at one time early in their evolution and one time near the end of hydrogen burning. Hydrogen mass fraction profiles are plotted to show how advanced the evolution is. The models shown are the same as in Figs.\,\ref{fig:vel-ev}b,d and have equal initial critical velocity fractions of 0.6. It can be seen that there is a point for each mass at which the specific angular momentum does not change.

Figure\,\ref{fig:angmomtran}.c shows this more clearly, where the difference in angular momentum at both times, divided by the angular momentum at the earlier time, $\Delta j/j_1$, is plotted. Here, regions where angular momentum is gained have a positive value, whereas regions were angular momentum is lost have a negative value. For the more massive model, the point with a constant specific angular momentum is closer to the star's edge than for the less massive model. For the 15\Msun model, approximately 90$\%$ of the total mass is acting as a donor of angular momentum, while for the 5\Msun mode the figure is 80$\%$. Using the arguments above, therefore the 15\Msun model will approach the critical rotation velocity more easily. Furthermore by inspecting the area under the curves, in the region where $\Delta j/j_1$ is positive in Fig.\,\ref{fig:angmomtran}.c one can determine how much relative angular momentum is gained. For example, if a region from $\frac{m}{M}=m_i$ to $\frac{m}{M}=1$ had doubled its total angular momentum, the integral $\int _{m_i} ^1 \Delta j/j_1 d(\frac{m}{M})$ would be equal to $(2-1)(1-m_i)$. Fig.\,\ref{fig:angmomtran}.c shows that the relative angular momentum gain of the matter in the envelope of the 5\Msun model is greater than that of the 15\Msun model. This tells us that to maintain near solid body rotation, a relatively smaller amount of angular momentum must be transported in the more massive model. 

\subsection{The Effect of Efficient Rotational Mixing \label{sec:CHE}}
When a star rotates initially at high velocities, quasi-chemically homogeneous evolution can occur. During such evolution, rotational mixing is so efficient that any chemical gradient between core and envelope cannot develop, meaning that the radiative envelope does not expand and the star's radius remains roughly constant \citep{1987A&A...178..159M,Yoon}. However because the luminosity of a quasi-chemically homogeneously evolving star approaches the Eddington limit, the critical velocity of such a star does decrease (through Eq.\,\ref{Eq:Vcrit}). Furthermore the increased luminosity causes a strong increase in the mass-loss rate, meaning that the equatorial velocities of stars undergoing quasi-chemically homogeneous evolution are likely to decrease with time. Thus such stars will evolve with a decreasing critical velocity fraction while quasi-chemically homogeneous evolution occurs.

This behaviour is shown for 25\Msun models in Figs.\,\ref{fig:vel-ev} e,f. It can be seen that the highest critical velocity fractions occur during the early part of the stars' lifetimes because the critical velocities (dashed lines) decrease relatively slowly while the equatorial velocities (dot-dashed lines) fall due to angular momentum loss. In the MW model, quasi-chemically homogeneous evolution is shutdown when the star reaches an age of around 80$\%$ of the hydrogen burning lifetime and from this point on the star evolves normally and advances towards the critical velocity. The phase of quasi-chemically homogeneous evolution ends because as the rotational velocity decreases, rotational mixing processes become less efficient and a chemical gradient eventually develops in the star which provides a barrier to mixing through buoyancy forces \citep{2000ApJ...528..368H} and effectively "turns off" quasi-chemically homogeneous evolution. 

For a star to evolve to a high critical velocity fraction, it must have a significant initial rotation velocity but also be rotating slowly enough to avoid quasi-chemically homogeneous evolution. As the minimum rotation rate required for quasi-chemically homogeneous evolution decreases with increasing mass \citep{Yoon}, very massive stars rotate at high critical velocity fractions for only very short fractions of their lifetimes, thus explaining the rarity of Oe stars.

\subsection{Nitrogen Enrichment}
\label{sec:N}
\begin{figure*}
	\includegraphics[width=0.9\linewidth]{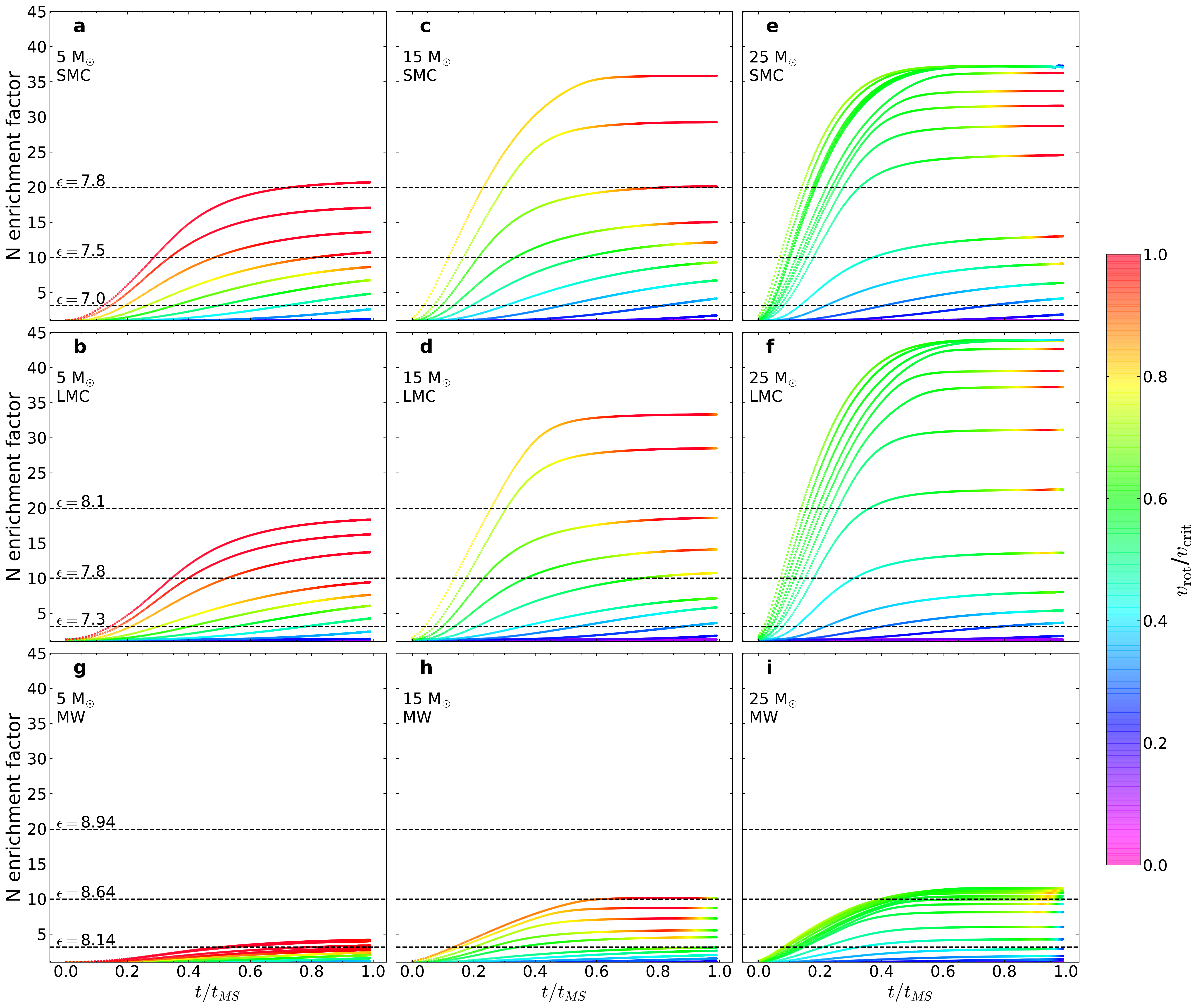}
	\centering
	\caption{ The surface nitrogen enrichment factor, computed as the nitrogen mass fraction divided by the initial nitrogen mass fraction as a function of the fractional hydrogen-burning lifetime, $t/t_{MS}$ for models with initial rotational velocities between 0 and 600\kms and initial masses 5, 15, 25\Msun as marked in the plot. SMC, LMC and MW compositions are shown in the top, middle and bottom panels respectively. The colour scale corresponds to the critical rotation fraction, \Vcritfrac. For each metallicity, various nitrogen abundances, $\epsilon = 12 + \textrm{log}(N/H)$ are displayed as dotted black lines with the value given in the left column plots.  }
	\label{fig:N} 
\end{figure*}

Here we address the question of whether or not Be stars formed through a single star evolving towards the critical velocity are expected to show significant surface nitrogen enrichment, where nitrogen is the product of hydrogen burning and is brought to the stellar surface through rotationally induced mixing. Fig.\,\ref{fig:N} shows the evolution of both surface nitrogen abundance and critical velocity fraction as a function of the fractional hydrogen-burning lifetime. Displayed are models with initial masses 5, 15 and 25\Msun with SMC, LMC and MW metallicities. As discussed by \citet{Brott1} the relative increase of the nitrogen abundance goes down  with increasing metallicity, therefore we see weaker nitrogen enrichment in the MW models than the LMC or SMC models. It is also clear that rotationally induced mixing is more efficient in more massive stars, owing to the effects of increased radiation pressure in more massive stars \citep{1987A&A...178..159M,Yoon}.

From the bottom panels of Fig.\,\ref{fig:N}, we expect that in the Milky-way, nitrogen is never enhanced by much more than a factor of 10 for models that rotate near the critical velocity. On the other hand, the LMC and SMC models that attain near critical rotation velocities show surface nitrogen enhancements of at least a factor 10 and up to approximately a factor 30.

We therefore judge that single Be stars in the LMC should have surface nitrogen abundances $\epsilon = 12 + \textrm{log}(N/H)$ no smaller than around 7.7, and in the SMC no smaller than around 7.4. In the Milky-way, we no not expect the single Be stars to have outstanding nitrogen surface abundances.

\section{Population Synthesis Results}
\label{sec:popsyn}

\subsection{Predicted Fractions of Be Stars}
\label{sec:Befrac}

\begin{figure*}
	\includegraphics[width=0.8\linewidth]{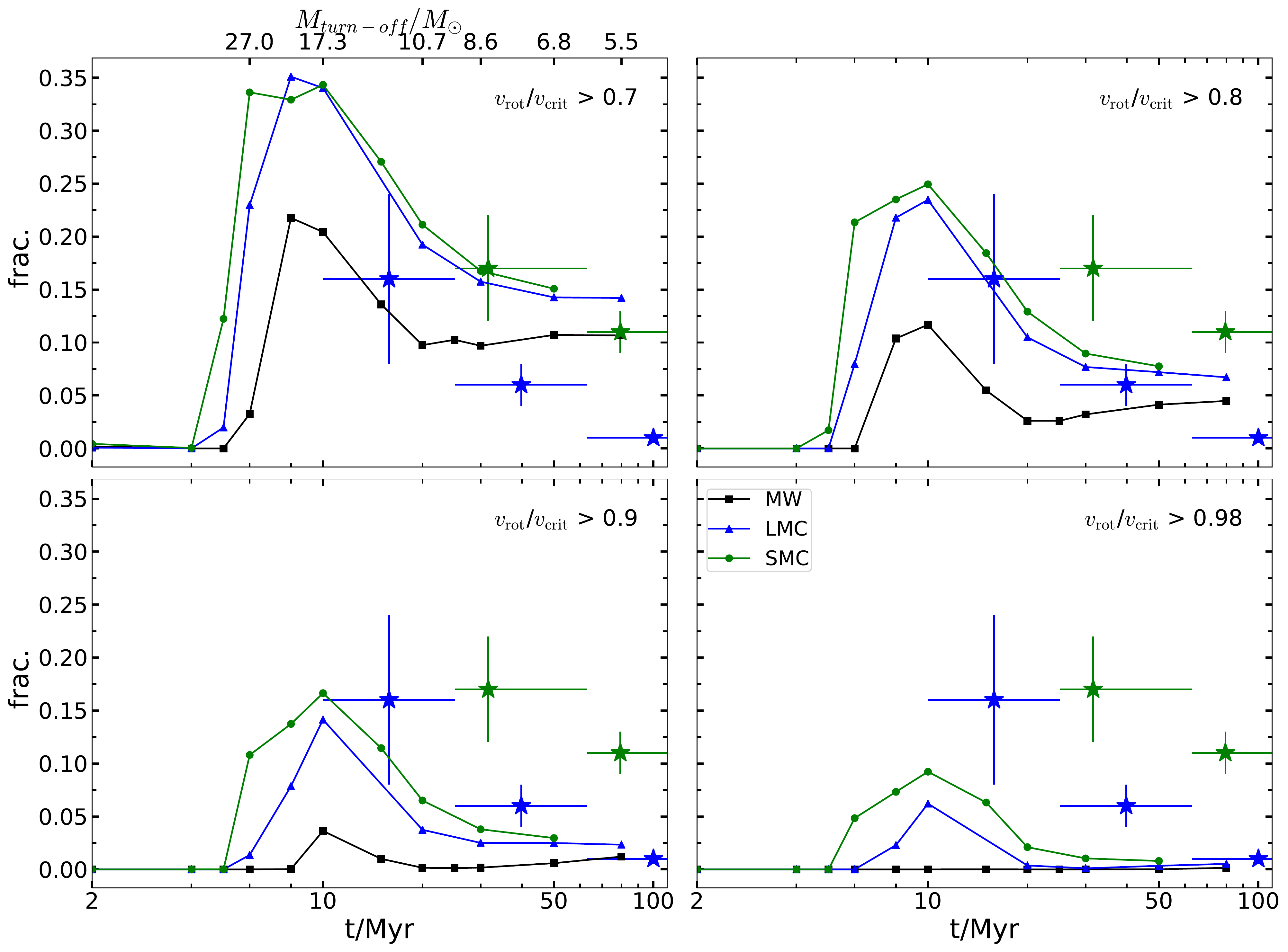}
	\centering
	\caption{The fractions of stars rotating faster than various values of the critical rotation fraction, \Vcritfrac, as given in the top left corner of each plot, in our synthetic coeval single star populations as function of their age. 
Considered only are stars brighter than one bolometric magnitude below the main-sequence turn-off. Metallicities are displayed as MW (black), LMC (blue) and SMC (green). For the top left panel the top scale gives the main-sequence turn off mass for non-rotating MW models. Observations with error bars from \citet{IqbalObs} are shown as green and blue stars for SMC and LMC observations respectively. }
	\label{fig:befrac} 
\end{figure*}

In this section, we discuss synthetic populations of coeval rotating single stars as described in Sec.\,\ref{sec:popsynmeth}.
These results can then be compared to the number of Be stars observed in young star clusters of various ages.  
From our models, we derive the fraction of Be stars within one bolometric magnitude 
(assumed to be equal to one visual magnitude) of the turn-off. We consider a stellar model to correspond to a
Be star when its rotational velocity exceeds a predefined fraction of critical rotation. 
Figure\,\ref{fig:befrac} shows the result as a function of age for various threshold critical velocity fractions and metallicities.

A striking feature of this plot is the maximum in Be fraction for all metallicities near 10\Myr. At $t=0$ there are no Be stars because, as discussed earlier, the initial rotation distribution prevents O stars entering the ZAMS with \Vcritfrac fractions greater than around 0.7. The 30\Msun models take approximately 5\Myr to evolve towards critical rotation, shortly after which point they leave the main-sequence. From 5 to 10\Myr the Be fraction grows sharply as angular momentum loss from winds diminishes. From 10 to 20\Myr the Be fraction falls because the of the core-mass effect as discussed in Section \ref{sec:coremass}. Comparing the hydrogen burning lifetimes of non-rotating MW models, it is found that populations with ages from 10 to 20\Myr have a main sequence turn off mass of around 17\Msun. These models are in a "Goldilocks" situation where they are massive enough to have an appreciable convective core but not so massive to lose large amounts of angular momentum. 

Furthermore we see that the Be fraction increases at earlier times for the lower metallicity models and that there is a clear trend in metallicity which shows that single Be stars become more common with decreasing metallicity. Both of these features are a result of the fact that the strength of angular momentum loss is metallicity dependant as discussed in Section \ref{sec:amLoss}.

It is also found that the Be fraction is strongly dependant on the chosen Be criterion. When critical velocity fraction, \Vcritfrac of 0.7 is chosen as the Be criterion, we predict Be fractions in the Magellanic Clouds in the range 15 to 35$\%$. Wheres when we restrict Be stars to being nearly critical rotators ( \Vcritfrac $> 0.98$), the Magellanic Cloud Be fraction lies in the range 0 to 10$\%$ and is almost 0 for population ages greater than 50\Myr. 

\subsection{Fast Rotators in the Colour-Magnitude Diagram}
\label{sec:HRdiagrams}

\begin{figure*} 
	\includegraphics[width=0.95\linewidth]{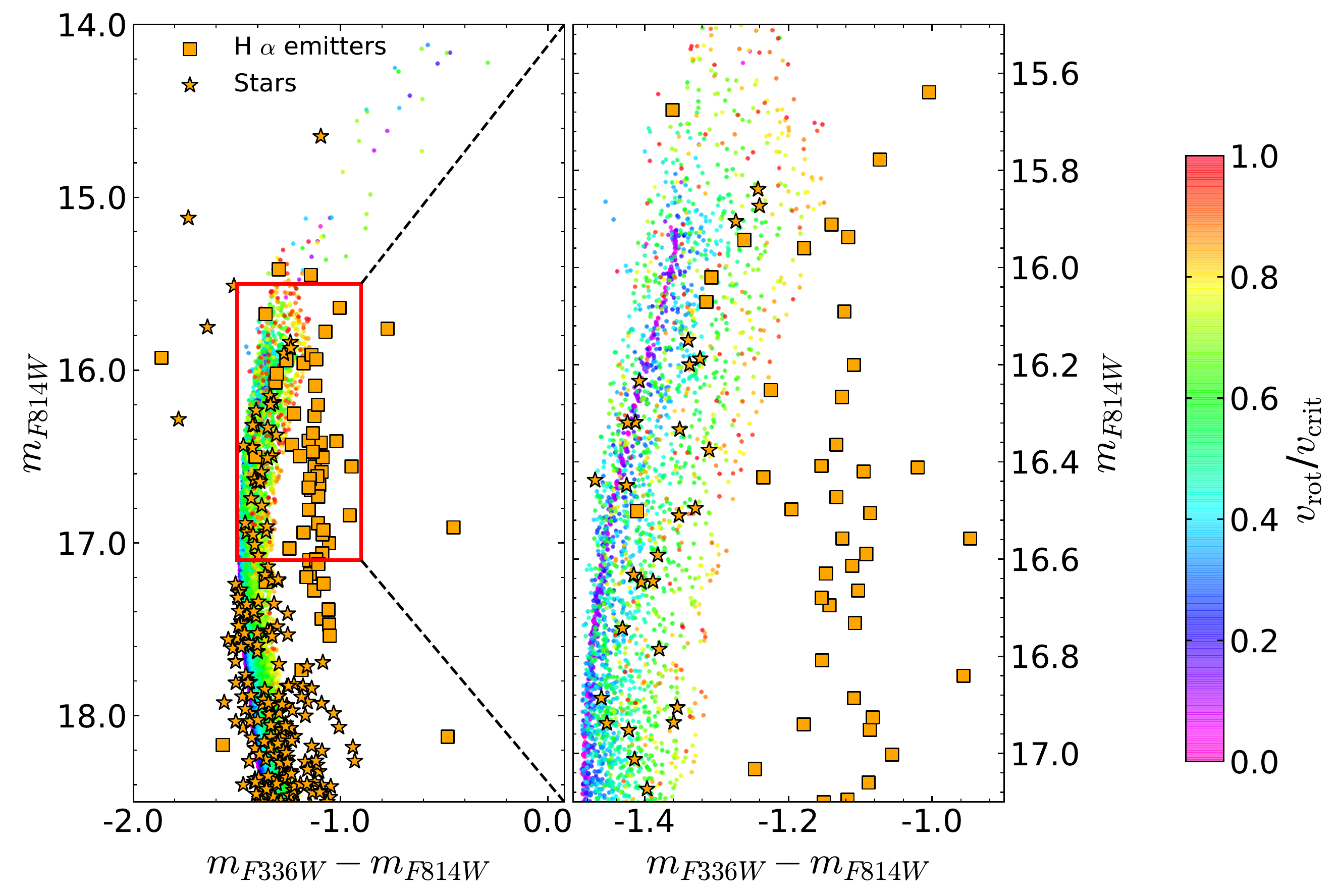}
	\centering
	\caption{ The synthetic colour-magnitude diagram of a 35Myr star cluster at SMC metallicity, where each dot represents one single star and the colour gives the critical velocity fraction, \Vcritfrac as indicated by the colour bar. Gravity darkening is included assuming a random orientation of the rotation axis. Over plotted are observations of SMC cluster NGC 330 \citep{MiloneObs2}, with H $\alpha$ emitters marked by orange squares and normal stars as orange star symbols. The right panel shows the region indicated by the red box in the left panel. To convert the models to apparent magnitudes a distance modulus of 18.92 mag and a reddening of 0.06 mag were used.}
	\label{fig:SMCHR} 
\end{figure*}

Using the procedure outlined in Section \ref{sec:popsynmeth} we build synthetic colour-magnitude diagrams to indicate the expected positions of fast rotators. Fig.\,\ref{fig:SMCHR} shows the colour-magnitude diagram positions and critical velocity fractions of our SMC models with a coeval age of 35\Myr. Over plotted on our theoretical predictions are Hubble Space Telescope observations of the SMC cluster NGC 330 \citep{MiloneObs}, 
We see that the nearly critically rotating stars are located very close to the turn-off, as can be expected by the fact that our models only achieve high fractions of critical rotation near core hydrogen exhaustion (see Figs.\,\ref{fig:vel-ev},\ref{fig:app1}). Fig.\,\ref{fig:SMCHR} also shows the effects of gravity darkening, with the slowly rotating models almost confined to a single isochrone while the fast rotators suffer strong gravity darkening and display a wider range of colours, due to a relatively large spread in effective temperatures. Appendix \ref{app:b} shows the same colour-magnitude diagram but ignoring the effects of gravity darkening for comparison.

In the following analysis we shall assume that the H$\alpha$ emitters in NGC\,330 are Be stars.
The observations in Fig.\,\ref{fig:SMCHR} show that most of the observed H$\alpha$ emitters are redder than the ordinary main sequence stars. 
This segregation is not a unique feature to NGC 330, with many LMC and SMC clusters exhibiting the same trait \citep{MiloneObs2}.
\citet{1998MNRAS.296..785T} have suggested through observations of Be stars whose spectra show rapid switching between containing 
emission lines and not, that the decretion disc can contribute up to 40$\%$ of a Be star's flux.
Because our models do not include the contribution of a Be star's decretion disc to the observed fluxes, it is not meaningful 
to compare the colours of our fast rotating models with those of observed Be stars. However, 
assuming that the error on the magnitudes of our synthetic Be stars is at most 0.35\,mag,
we may consider the relationship between the relative number of Be stars and apparent magnitude. 

\begin{figure} 
	\includegraphics[width=0.95\linewidth]{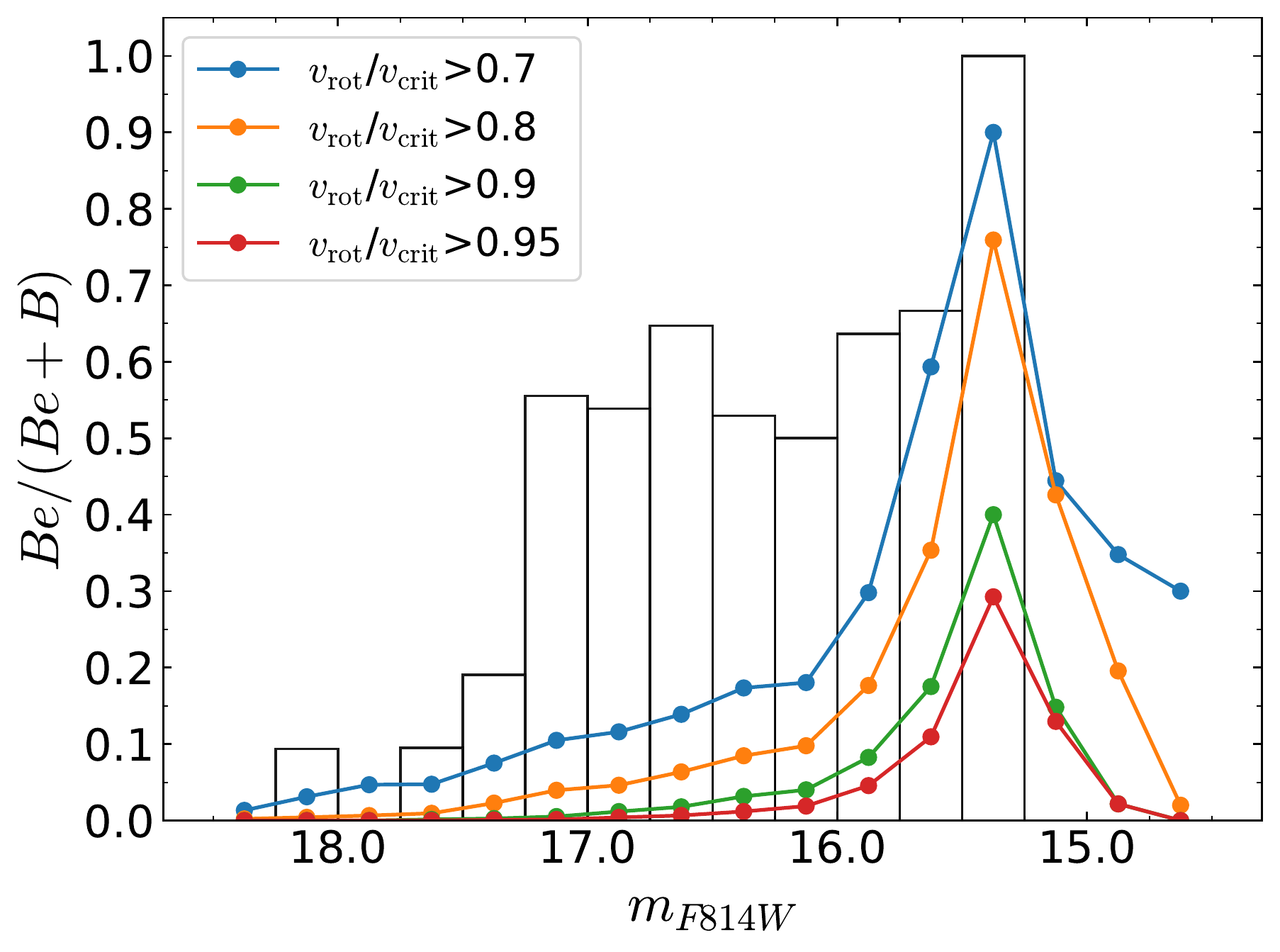}
	\centering
	\caption{ A histogram of the Be fraction of stars in SMC cluster NGC 330 \citep{MiloneObs2} as a function of apparent magnitude in the Hubble Space Telescope Wide-field Camera 3 filter F814W, $m_{F814W}$ plotted as white columns. It is assumed that H $\alpha$ emitters are Be stars. Over plotted as coloured lines are our model predictions in the same magnitude bins for various minimum rotation rates to be counted as a Be star, as given in the legend.}
	\label{fig:SMCHR_hist} 
\end{figure}

Fig.\,\ref{fig:SMCHR_hist} shows the Be fraction of our model predictions in F814W apparent magnitude bins for 
various threshold rotation rates for stellar models to be considered a Be star. We see that our model Be stars are strongly biased 
to being located near the main-sequence turn off, around $m_{F814W} = 15.4\,$mag, where depending on how fast we require a Be star to rotate, 
the Be fraction is between 30 and 90$\%$. Fig.\,\ref{fig:SMCHR_hist} compares our model predictions to Be star counts 
in NGC 330 \citep{MiloneObs2}. 
The observations show that in NGC\,330 the Be fraction is approximately 0.6, and it remains
rather constant within a wide magnitude range of 17.0 to 15.6\,mag. This result is quantitatively confirmed
by a recent study of \citet{Bodensteiner2020}, who used MUSE spectroscopy to identify Be stars
in the core of NGC\,330. 

Comparing our models with the observations, Fig.\,\ref{fig:SMCHR_hist} reveals that when assuming that Be stars are 
required to rotate only at 70$\%$ or more of the critical rotation velocity, 
our single star models agree with the observed Be star count in NGC\,330 at the turn off region.
However, they strikingly fail in two respects. Firstly, even adopting the least stringent threshold value for Be stars,
our models fail to produce the large number of observed Be stars. Secondly, our models can not 
reproduce the fact that the Be star fraction in NGC\,330 is constant over a range of 1.5\,magnitudes. 
  
Here, the second failure seems the worst. The total number of Be star can in principle be
boosted by lowering the rotation threshold for considering the models a Be star, or by adopting larger initial 
rotation velocities. However, it is an intrinsic feature of the rotating single star models
to increase the ratio of rotation to critical rotation velocity with time (Fig.\,5; see als Ekstr\"om et al. 2008) . 
Thus, it appears quite unlikely that single star evolution that the observed distribution of Be\,stars 
in NGC\,330 can be explained solely by single star evolution.

\section{Discussion}
\subsection{Uncertainties}
It is important to keep in mind that models are simply that and at some point they must fail to reflect the behaviour of real stars. \citet{Brottlifetimes} have found that the hydrogen burning lifetimes given by models studied here differ to those of MESA models by approximately 15$\%$. As this is merely a discrepancy in the clock, but not in the physical behaviour of the models \citep{pabloPhd}, this is concerning, but should not change the main results presented here.
 
Another issue is the treatment of mass-loss, which is a strong factor in determining a model's evolution towards the critical velocity. The mass-loss prescription used \citep{VinkMassloss} was calibrated for models in the range 15-20 \Msun , so it may be questionable whether this scheme is accurate for models outside this range. Furthermore the correct treatment of a star rotating close to the critical velocity is complex. Near the critical velocity, two distinct winds are expected to form, a cold equatorial wind (which carries away angular momentum) and a warm polar wind (which carries away less angular momentum). It is not apparent which wind has the dominating effect, although the winds of models presented here always carry away angular momentum. Eq.\,\ref{Eq:massloss} demands that as a star approaches the critical rotation velocity, the mass loss rate becomes infinite. It can also be questioned whether this is a correct treatment of the mass loss of a critically rotating star.

In the SMC, \citet{2018MNRAS.476.3555R} matched light-curve models to observations of 54 Be stars and determined that the typical mass-loss rate of a Be star in the mass range 10 to 20\Msun is of the order $10^{-10}$\Msun yr$^{-1}$. Our models predict that only stars with initial masses less than around 10\Msun have such mass-loss rates on the main-sequence. This discrepancy highlights the fact that the models studied here are 1-dimensional and hence may struggle to represent accurately the mass-loss of a rotating star (which is a 2-dimensional problem).   

Our definition of a Be star is one that is rotating close to the critical velocity. However it would appear that nature has a slightly different definition, with pulsations perhaps playing a role  \citep{2001A&A...369.1058R,2014IAUS..301..465N}. Pulsations could serve to kick matter off of the stellar surface and aid the formation of a circumstellar disc. Observations with the CoRoT space telescope show that Be stars display pulsations that can transport angular momentum through the star \citep{2009A&A...506...95H}, thus affecting the evolution of rotation velocities. The interaction of rotation and pulsations is out of the scope of this work and the results of such endeavours are eagerly awaited. Furthermore recent observations with the TESS space telescope imply that the Be star disc could be fed by mass ejections from starspots \citep{2019arXiv191103068B}. Be stars are complex objects, with rotation being a key ingredient of the Be phenomenon, but perhaps not the only one.

Our population synthesis results are dependant on the initial rotation distribution that is assumed. Whereas \citet{VFTSB}
used high quality data, a large and unbiased sample of star, and corrected for effects as macro turbulence, other rotational
velocity distributions are available \citep{2006A&A...452..273M,2008A&A...479..541H,HuangGiesObs}
While our quantitative results might change using one of them, we do not expect a change in the qualitative behaviour of our results,
which appears to be determined by the evolutionary factors discussed in Sect.\,\ref{sec:results}

\subsection{Comparison with Previous Models}

The frequencies of single Be stars have been predicted from models by \citet{EkstromBe}, \citet{GranadaBe} and \citet{GranadaBe2}. A major difference between these sets of models and the models studied here is the inclusion of the effects of an internal magnetic field which strongly couples the core and envelope, which may increase the predicted numbers of critically rotating stars. 

\citet{EkstromBe} predict at solar metallicity and an age of 25\Myr the fraction of stars with a brightness of up to two magnitudes below the turn-off and rotating at the critical velocity is 5$\%$, compared to 0$\%$ found here. However when one looks at the fraction of stars rotating faster than \Vcritfrac =0.7 at 20\Myr, \citet{EkstromBe} finds a Be fraction of around 15$\%$, compared to 8$\%$ found here. Both sets of models agree that at ages greater than around 40\Myr, almost no stars rotate at the critical velocity.
\citet{EkstromBe} used a Gaussian-like initial rotation distribution with a peak at $\Omega / \Omega_{\textrm{crit}} = 0.6$, which in the Roche model corresponds to \Vcritfrac $\approx$ 0.4, and therefore is judged to be a similar initial rotation distribution to the one used in this work.   
As demonstrated by \citet{EkstromBe} mass loss rates play a crucial role in the evolution of the surface rotation. In our models above 10\Msun , a turn over in \Vcritfrac is caused by strong mass-loss at late times (see Figure \ref{fig:vel-ev}). When using the \citet{KudritzkiMassloss} mass-loss scheme, \citet{EkstromBe} found this turn over not to occur, and so those models spend more time at high \Vcritfrac values and so Be stars become more common. It is therefore concluded that mass-loss is just as important as angular momentum transport in producing stars which rotate close to the critical velocity at galactic metallicity. At $Z=0.002$ (a metallicity similar to our SMC models) \citet{EkstromBe} calculate a maximum in Be fraction at an age of 10Myr of around 10$\%$, in good agreement with the results presented here.

Both \citet{EkstromBe} and \citet{GranadaBe} find that Be stars should become rarer at lower metallicities, which is in contradiction to the results presented here and the general trend seen by observers \citep{IqbalObs,MaederObs,MatayanObs}.
 
\subsection{Comparison with Further Observations}

\citet{IqbalObs} observed Be star fractions within 1 visual magnitude of the turnoff in LMC and SMC clusters. Such observations are directly comparable with Fig.\,\ref{fig:befrac}, with the data over-plotted on our predictions. In clusters of ages from 7 to 8 Myr the Be fraction decreases from 15 $\%$  to 0 $\%$. When defining the Be criterion to be 0.9 \Vcritfrac, a similar behaviour is found albeit at later ages. \citet{IqbalObs} also find that Be fractions increase with decreasing metallicity, in agreement with our models. 

Both \citet{MaederObs} and \citet{MatayanObs} found that Be stars are three to five times more frequent in the SMC than the galaxy, again in fairly good agreement with the model predictions. \citet{MatayanObs} reported that the distribution of Be star frequency across spectral types does not depend on metallicity. As Fig.\,\ref{fig:befrac} shows similar trends for all metallicities, this behaviour is confirmed by the models. 

Observations from \citet{TarasovObs} show that Be stars become most common in clusters with ages of 12-20Myr, in relatively good agreement to the model predictions.

\citet{OeSMCObs} found that in the SMC the frequency of Oe stars is strongly peaked around spectral types O9. Furthermore the Oe to O star fraction was measured as 0.26, compared to 0.03 for the MW. This measurement supports our result that very few stars with high critical velocity fractions and ages less than 10Myr should be found in the MW, but are found at lower metallicities (see lower Panels of Fig.\,\ref{fig:befrac}).

Owing to the fact that the results of our population synthesis rely strongly on the adopted initial rotation distribution, one may question whether it is appropriate to assume that stars in the MW and Magellanic Clouds have equivalent initial critical rotation fractions. Whether the observed trends in Be fraction with metallicity are due to stellar evolution or a metallicity dependant rotation distribution is not clear. If lower metallicity stars were to rotate significantly faster, rotationally enhanced mass-loss would hinder the formation of Be stars at lower metallicities, therefore there is a limit to how much faster stars at lower metallicities can rotate. Even though \citet{Penny} found no significant difference between rotational velocities of O-type stars in the galaxy and Magellanic Clouds, \citet{Keller} concludes that LMC stars are more rapidly rotating than galactic stars. It is curious to consider that even if LMC stars have faster equatorial velocities, due to their compactness they have larger critical velocities, and hence perhaps the same initial \Vcritfrac as galactic stars.

\subsection{Comparing the Single and Binary Star Formation Channels of Be Sstars.}

Whereas in this work we focus on the single star formation channel for Be stars, it is evident that
Be stars can also be formed through close binary evolution. The main mechanism is
spin-up by accretion, which is expected to occur as a consequence of mass transfer (Langer 2012).
The class of Be/X-ray binaries \citep{2011Ap&SS.332....1R} provides strong support for this picture.
\citet{PolsBinaryModels} showed through detailed models that Be stars may be 
produced by mass accretion from a companion star in the course of close binary evolution. Using simplified binary
evolution calculations, \citet{2014ApJ...796...37S} demonstrated that potentially a large enough number of them
could emerge from binary evolution to explain the currently observed Be star populations. 
In the following, we discuss several basic differences which can be expected between Be stars formed through binary interaction 
compared to those formed through the single star channel. 

As we have seen above, when using the rotational velocity distribution of \citet{VFTSB}, 
it is difficult for single star models to achieve very close to critical rotation
(lower right panel of Fig.\,\ref{fig:befrac}). To remedy this would require that a significant fraction of stars is 
already born with near critical rotation. For low enough mass or metallicity, this group of stars would the remain rapidly rotating
throughout their main sequence life. Such a picture appears not to be supported by observations \citep{2005ApJS..161..118M}.
Accordingly, single star evolution appears to be able to explain significant Be star populations only if decretion discs could form 
also in stars rotating significantly below critical. 

In mass transferring binary evolution models, this is different. The angular-momentum of the mass-gainer increases quickly, 
such that critical rotation can be achieved after a relative mass increase of the mass gainer of 10\% or less
\citep{1981A&A...102...17P}. During mass-transfer there is no fine-tuning mechanism that switches off accretion 
when a given rotation rate is reached, instead the only limit is critical rotation.
Therefore, all mass transferring binaries where tides do not limit the spin-up of
the mass gainer --- which is the vast majority --- are expected to produce a critically rotating main sequence star
\citep{2012ARA&A..50..107L}. After the accretion phase, the two mechanisms which affect the single stars, i.e., spin-down by mass loss and spin-up
due to core contraction, will also work in the spun-up mass gainer. Whereas the wind induced drain of angular momentum
may spin down some of the most massive mass gainers, the core contraction accompanying central hydrogen burning ensures
that most of them remain at critical rotation for the rest of their main sequence evolution. 

Consequently, whereas single star evolution leads to an increase of the rotation velocity compared to 
its critical value in many cases, the binary channel can produce a much larger number of stars living at critical rotation
for a long time, compared to the single star channel.
Furthermore, \citet{Chen2020} find that the initial 
mass ratio limit for stable mass transfer leads naturally to a restriction of binary-produced Be stars 
to within about two magnitudes of the cluster turn-off, which compares well with observations when 
interpreting the H$\alpha$ emitting stars in NGC\,330 as Be stars \citep{MiloneObs2}. 

A further important difference between the Be stars produced via single and binary evolution concerns their
expected surface abundances. As discussed above, the mass-gainer of a binary system may only accrete a small amount of mass 
to spin up. In this case, only material from the outer envelope of the donor star is incorporated into the mass gainer. As this material is 
generally not enriched in hydrogen burning products, one would expect that Be stars formed through the binary channel 
are not significantly polluted by accretion. Detailed binary evolution models with LMC metallicity \citep{Langer2019} suggest that 
the surface nitrogen mass fraction of spun-up mass gainers is at most tripled compared to the baseline nitrogen abundance.
Additionally, the spun-up mass gainers had ordinary rotation rates before the mass transfer episode. As such, they established
a strong mean molecular weight barrier between core and envelope, which prevents any significant rotationally induced mixing
after their spin-up. 

In Sec.\,\ref{sec:N} we argue that single Be stars ought to have much larger surface nitrogen enhancements (by as much as a factor 30;
cf., Fig.\,\ref{fig:N}). A diagnostic to discriminate single and binary Be stars would then be their surface nitrogen abundances. 
\citet{BeN1} find that in the LMC cluster NGC 2004, only two Be stars from a sample of 11 were measured to have a 
nitrogen abundance $\epsilon = 12 + \textrm{log}(N/H)$ greater than 7.8, while the other 9 Be stars had $\epsilon$ less than 7.4. 
This observed bimodal distribution supports the idea of the binary and single Be star formation channels producing populations 
with different nitrogen enrichments, and would suggest that in NGC 2004 the binary formation channel dominates. 
Also \citet{BeN1} found that the nitrogen abundances of the majority of the LMC Be stars observed in the 
VLT-FLAMES Survey of Massive Stars are not consistent with single star evolution.  
Also, the Be star NGC 330-B 12 was found to be almost devoid of nitrogen lines and possessing a spectrum inconsistent 
with single star evolution models \citep{2005A&A...438..265L}, giving further evidence that Be stars can be formed by binary interactions.

Finally we note that a key difference between Be stars produced by the two channels concerns their potential binary companions.
Since the initial rotational velocity distribution for single stars and stars in binaries appears to be similar
\citep{2015A&A...580A..92R}, we would expect a significant fraction of Be stars formed through the single star
channel (i.e., without accretion-induced spin-up) to have unevolved main sequence companions. However, essentially no such stars
are known. Vice versa, whereas massive binary-produced Be stars may be single since they lost their companion
when it produced a supernova explosion, the lower mass binary-produced Be stars should all have evolved companions:
subdwarfs or white dwarfs. While those are very hard to detect \citep{2018A&A...615A..30S},
recent studies of disc truncation of apparently single Be stars suggest that indeed unseen companions are present
in the majority of cases \citep{2019arXiv190912413K}.

\section{Conclusions}

We have identified and discussed three factors which affect a star's evolution towards the critical velocity 
throughout main-sequence evolution. Mass-loss through stellar winds has the effect of removing angular momentum from a star, 
and so hinders the approach to the critical velocity. The fraction of convective core mass to the total stellar mass strongly 
affects the internal angular momentum transport, which is crucial for an expanding envelope to maintain a fast rotational velocity. 
Lastly the occurrence of quasi-chemically homogeneous evolution prevents the stellar envelope from expanding and 
thus critical velocity decreasing, and also increases the angular momentum lost through stellar winds. 

When using an observed distribution of B\,star rotational velocities for constructing synthetic stellar populations, 
we find that our single star models predict few stars rotating at near critical velocities, 
although we do predict as much as 35$\%$ of OB stars to rotate with more than 70$\%$ of their critical velocity. 
We therefore conclude that if Be stars are near-critical rotators, then single star models cannot explain the observed numbers of Be stars. 
In this case, most Be stars must be the product of mass-transfer in binary systems. 

If Be stars instead only rotate at say 70-80$\%$ of their critical velocity, then the observed Be star fractions 
can be reasonably described by single star evolution (see Fig\, \ref{fig:befrac}). However, in the $\sim$40\Myr old SMC cluster NGC\,330, 
Be stars are observed in significant numbers down to almost two magnitudes below the main-sequence turn-off. 
Independent of the rotation threshold for the Be phenomenon, our single star models predict that Be stars 
should be located only in a narrow luminosity range near the turn-off (see Fig\, \ref{fig:SMCHR_hist}), which disagrees with observations of NGC 330.

Whereas significant uncertainties remain, specifically in reconciling how stars that appear to rotate at 70$\%$ of their critical velocity can still form decretion discs and why so few Be stars are observed to rotate near the critical velocity, it appears evident that the observed Be star populations can not be explained
by single star evolution alone, and that it may not be the dominant channel for Be star formation. Nevertheless,
single star evolution will contribute, most strongly so in the age range from 8 to 20\,Myr, at least at sub-solar metallicity.

Furthermore, our single star models predict that the surfaces of rapidly rotating single stars 
should be contaminated with freshly synthesised nitrogen, the more the faster the rotation.
In spun-up mass gainers of binary systems, this is not necessarily so. The observations of non- or weakly
nitrogen enriched surfaces in several groups of Be stars therefore strengthens the conclusion that
the majority of these objects can not originate from single star evolution.

\bigbreak
Acknowledgements. We are grateful to Dietrich Baade, Danny Lennon and Christophe Martayan for discussions and for pointing out the recent TESS result to us. We also thank our anonymous referee for useful comments on an earlier version of this manuscript.

\bibliographystyle{aa_url} 
\setlength{\bibsep}{0pt}
\bibliography{bib}{}

\clearpage

\onecolumn

\begin{appendix}
\section{Initial Conditions Required to Reach Near Critical Velocities \label{app:a}}
\setcounter{figure}{0}

\begin{figure*} 
	\includegraphics[width=0.98\linewidth]{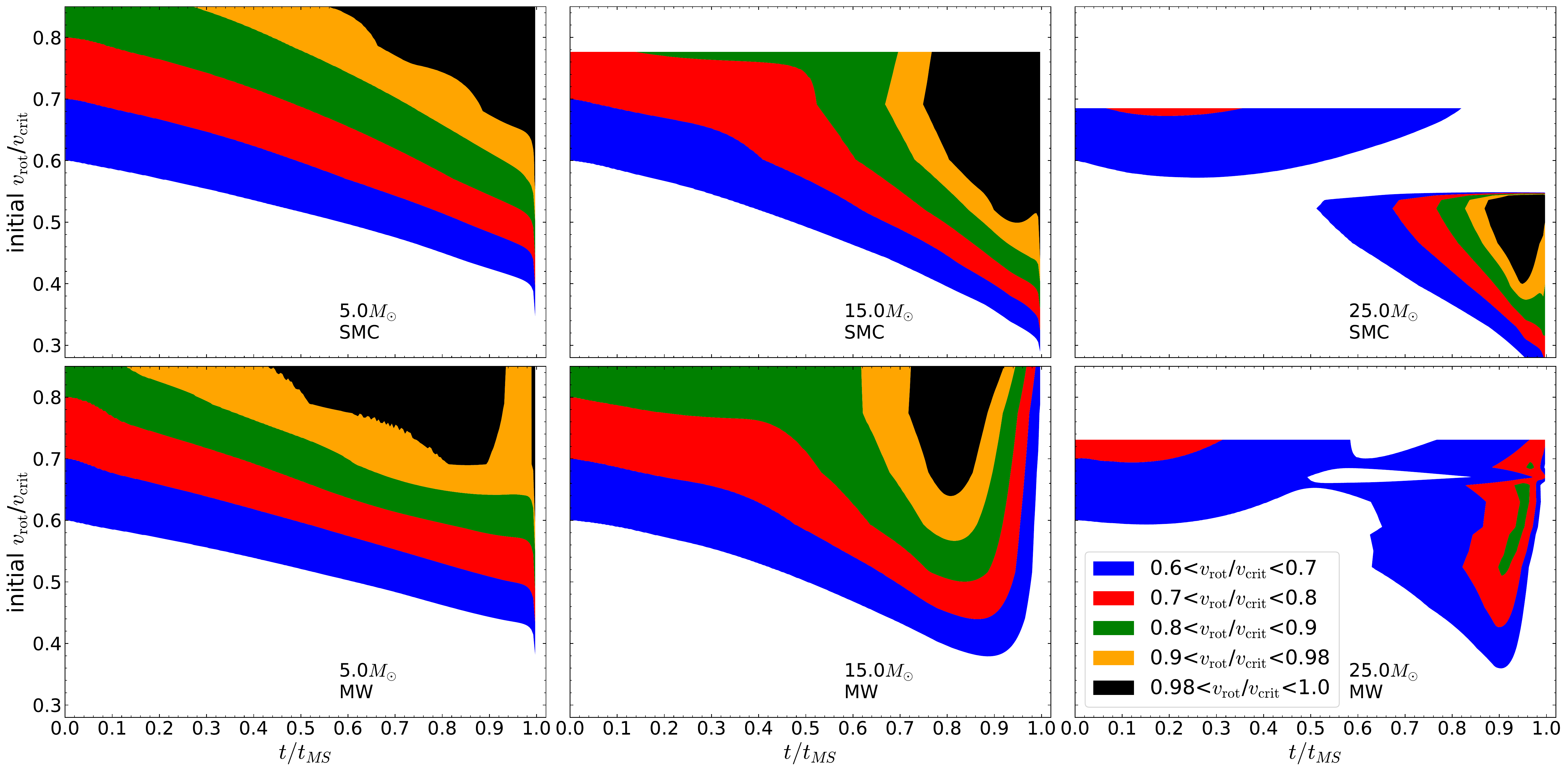}
	\centering
	\caption{The critical velocity fractions, \Vcritfrac as a function of fractional hydrogen-burning lifetime, $t/t_{MS}$ and initial critical velocity fractions for models with masses of 5, 15 and 25\Msun and SMC (top panels) and MW (bottom panels) metallicities, as indicated in the figure. The colours indicate the critical velocity fraction, \Vcritfrac as given in the legend.  }
	\label{fig:app1} 
\end{figure*}

In Section \ref{sec: velEv} we have discussed the evolution of rotational velocities of various models, when the models all have the same relative initial rotation rates. Here we explore how the approach to the critical velocity depends on the initial rotation rate by performing interpolations between the models. Fig.\,\ref{fig:app1} shows, for differing stellar masses and metallicities, the critical velocity fraction as a function of time and initial rotation rate. The colour of each point on the plot shows the critical velocity fraction at a particular fractional hydrogen-burning lifetime, $t/t_{MS}$, and initial critical velocity fraction value. By following horizontal lines in the plot, one traces the evolution of a single model through its evolution.

For 5\Msun models at both SMC and MW metallicities, the star evolves generally towards higher critical velocity fractions. This can be seen as one traces a horizontal line, one moves always into regimes of higher critical velocity fractions. The exception is the 5\Msun MW models with \Vcritfrac $>0.7$, which at the end of hydrogen burning spin down through increased rotationally enhanced mass-loss. Looking at 15\Msun models at SMC metallicity, one sees also that there is a constant evolution towards higher critical velocity fractions. On the other hand 15\Msun models with MW metallicity evolve to higher critical velocity fractions until around $80\%$ of the hydrogen-burning lifetime, then they spin down due to angular momentum loss through winds (as discussed in Sec.\,\ref{sec:amLoss}). The 25\Msun models behave in a more complicated way because the initially very fast rotating models can undergo quasi-chemically homogeneous evolution (as discussed in Sec.\,\ref{sec:CHE}). In the right panels of Fig.\,\ref{fig:app1} one can see the two regimes according to whether or not the critical velocity fraction is increasing or decreasing. We can see that for 25\Msun MW models with initial \Vcritfrac $\approx 0.7$, they evolve at first to lower critical velocity fractions then after $t/t_{MS} \approx 0.6$ they begin to evolve towards higher critical velocity fractions. This occurs because initially the star is evolving quasi-chemically homogeneously, during which time rotation rates and hence rotationally induced mixing efficiency drops until quasi-chemically homogeneous evolution is shut down, at which point the star begins to evolve with an expanding envelope and approaches the critical velocity. For similarly initially fast rotating SMC models the same behaviour does not occur due to the weaker stellar winds at lower metallicities.

Fig.\,\ref{fig:app1} also shows us that the models only reach critical rotation (the black areas in the figure) very near core hydrogen exhaustion. Furthermore, when one looks at the minimum initial rotation rate required to reach near critical rotation, it decreases with increasing mass due to angular momentum transport efficiency ( as discussed in Sec.\,\ref{sec:coremass}). 

\section{Synthetic Colour-magnitude Diagram Without Gravity Darkening \label{app:b}}

\begin{figure*} 
	\includegraphics[width=0.95\linewidth]{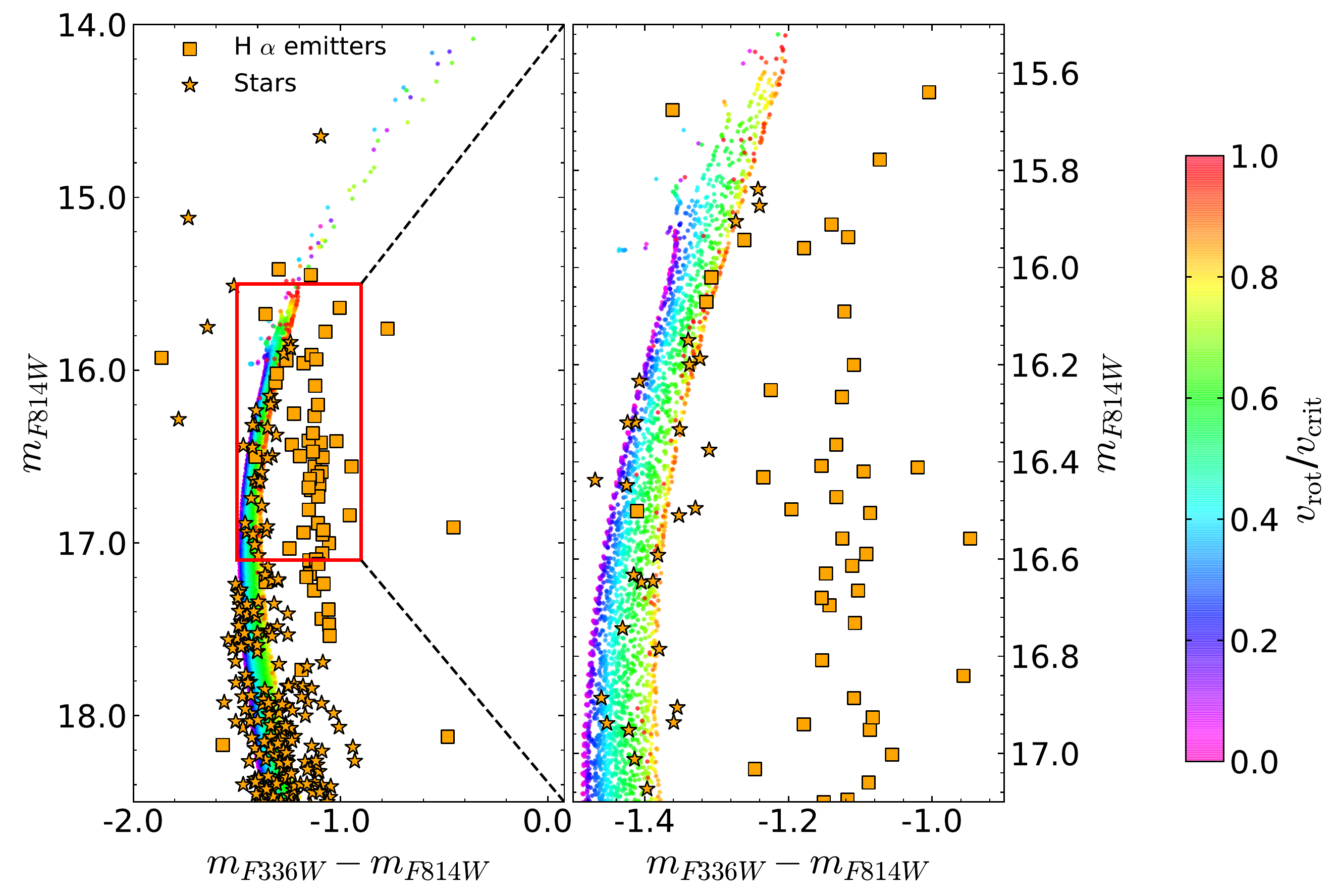}
	\centering
	\caption{ The synthetic colour-magnitude diagram of a 35Myr star cluster at SMC metallicity, where each dot represents one single star and the colour gives the critical velocity fraction, \Vcritfrac as indicated by the colour bar. Gravity darkening is not included. Over plotted are observations of SMC cluster NGC 330 \citep{MiloneObs2}, with H $\alpha$ emitters marked by orange squares and normal stars as orange star symbols. The right panel shows the region indicated by the red box in the left panel. To convert the models to apparent magnitudes a distance modulus of 18.92 mag and a reddening of 0.06 mag were used.}
	\label{fig:SMCHR_noGD} 
\end{figure*}

In Fig\,\ref{fig:SMCHR_noGD} we present the results of our population synthesis to simulate the colour-magnitude diagram of NGC 330 while ignoring the effects of gravity darkening. It is seen that along the main sequence there is a one-to-one relation between current critical velocity fraction and the $m_{F336W} - m_{F814W}$ colour, with faster rotators being redder. Such a relation is destroyed by gravity darkening (see Fig.\,\ref{fig:SMCHR}).

\end{appendix}

\end{document}